# Evidence of Low-Temperature Superparamagnetism in $Mn_3O_4$ Nanoparticle Ensembles

R. J. Tackett[1], J. G. Parsons[2], B. I. Machado[3], S. M. Gaytan[3], L. E. Murr[3], and C. E. Botez[1,a)]

[1]*Department of Physics, University of Texas at El Paso, 500 W. University Avenue, El Paso, TX 79968, U.S.A*

[2]*Department of Chemistry, University of Texas – Pan American, 1201 W. University Drive, Edinburg, TX 78539, U.S.A*

[31]*Department of Metallurgical and Materials Engineering, University of Texas at El Paso, 500 W. University Avenue, El Paso, TX 79968, U.S.A*

Using ac-susceptibility measurements and transmission electron microscopy, we have investigated the magnetic behavior of $Mn_3O_4$ nanoparticle ensembles at temperatures below the paramagnetic-to-ferrimagnetic transition of the title material ($T_N \cong 41K$). Our data show no evidence of the complex magnetic ordering exhibited by bulk $Mn_3O_4$. Instead, we find a low-temperature ($T<T_N$) magnetic anomaly that manifests itself as a peak in the out-of-phase component of the ac-susceptibility. Analysis of the frequency and average-particle-size dependence of the peak temperature demonstrates that this behavior is due to the onset of superparamagnetic relaxation, and not to a previously hinted at spin-glass-like transition. Indeed, the relative peak temperature variation per frequency decade $\Delta T/T\Delta \log(f)$ is 0.11, an order of magnitude larger than the value expected for collective spin freezing, but within the range of values observed for superparamagnetic blocking. Furthermore, attempts to fit the frequency f /observation time $\tau=1/2\pi f$ dependence of the peak temperature by a power law led to parameter values unexpected for a spin-glass transition. On the other hand, a Vogel-Fulcher law $\tau = \tau_0 \exp[E_B/k_B(T-T_0)]$ - where $E_B$ is the energy barrier to magnetization reversal, $k_B$ is Boltzman constant, $\tau_0$ and $T_0$ are constants related to the attempt frequency and the interparticle interaction strength - correctly describes the peak shift and yields values consistent with the superparamagnetic behavior of a slightly interacting system of nanoparticles ($\tau_o \sim 10^{-10}$s, $E_B/k_B=87K$, and $T_0=4.8K$). In addition, the peak temperature T is sensitive to minute changes in the average particle size $<D>$, and scales as $(T-T_0) \propto <D>^3$, another signature of superparamagnetic relaxation.

a) Corresponding author: E-mail: cbotez@utep.edu. Tel: 915-747-8040

**INTRODUCTION**

Investigations of the magnetic behavior exhibited by *bulk* $Mn_3O_4$ at temperatures below its paramagnetic-to-ferrimagnetic transition ($T_N \cong 41$ K) revealed the existence of complex (long-wavelength) ordering induced by two consecutive transitions: one to an incommensurate magnetic state at 39 K, and another to a commensurate state upon further cooling to 33 K [1,2]. More recent work confirmed these transitions, and also demonstrated that these low temperature magnetic states exhibit magnetodielectric coupling [3,4]. $Mn_3O_4$ *nanoparticles*, however, seem to show a very different magnetic behavior within the same temperature range. As reported by Regmi *et al.* [5], the above-mentioned magnetic ordering is suppressed in a 20-nm-average-size $Mn_3O_4$ nanoparticle ensemble. Instead, a single magnetic event is observed below $T_N$, namely a weak frequency-dependent peak in the out-of-phase component of the ac-susceptibility $\chi''$ at about 31 K. Interestingly, the microscopic origin of this magnetic behavior has not been fully elucidated. The authors of Ref. 5 hint at surface-spin-freezing effects, yet, as we argue below, the superparamagnetic relaxation of the nanoparticle ensemble could be a viable alternative explanation.

Due to the similarity of their basic ac-susceptibility signature – a peak in the $\chi''$ vs. T dependence – distinguishing between superparamagnetic blocking [6,7] and collective spin-glass-like freezing (of either surface spins [8,9] or superspins [10,11]) is often not straightforward. However, close examination of the frequency and average-particle-size dependence of the peak temperature may be used to differentiate between the two types of magnetic behavior [12]. For example, the relative variation of the peak temperature per frequency decade associated with a spin-glass-like transition is expected to be an order of magnitude smaller than that exhibited by the onset of superparamagnetic blocking [13-15]. In addition, the frequency/observation time dependence of the peak temperature for collective spin freezing follows the conventional dynamic scaling theory [16,17], while its counterpart associated with superparamagnetic blocking is best described by either the Néel-Brown equation (for ideal non-interacting ensembles)[18] or the phenomenological Vogel-Fulcher law [19,20] (for slightly interacting systems). Finally, the collective spin freezing temperature does not change significantly with the nanoparticle dimensions, while the superparamagnetic blocking temperature exhibits a pronounced dependence on the ensemble's average particle size [21].

Here we present a study aimed at clarifying the low-temperature magnetic behavior of $Mn_3O_4$ nanoparticle ensembles. Our temperature-resolved ac-susceptibility data collected on <D>=13 nm (average-size) nanoparticles at different frequencies f confirm the suppression of the long-wavelength magnetic ordering exhibited by bulk $Mn_3O_4$. The observed $\chi''$ vs.T dependences exhibit two peaks: one at 40 K, and another near 11 K. The first peak is frequency-independent, and corresponds to the well documented paramagnetic-to-ferrimagnetic transition of $Mn_3O_4$ [1-5]. The low-temperature peak, however, is frequency-dependent, and since it represents the sole magnetic anomaly below $T_N$, we believe it has same microscopic origin as its counterpart observed in larger nanoparticles (<D>=20 nm) at T~31 K (in Ref. [5]). Yet, such common microscopic origin is not likely to be collective spin freezing, as there is no reason for the freezing temperature to decrease from 31 K to 11 K upon the average nanoparticle size reduction from 20 nm to 13 nm. On the other hand, the blocking temperature associated with the superparamagnetic relaxation of a ferromagnetic or ferrimagnetic nanoparticle ensemble $T_B$ is predicted to depend on $<D>$. For an ideal system of magnetic nanoparticles, for example, $T_B \propto \dfrac{K<D>^3}{k_B}$, where K is the magnetic anisotropy and $k_B$ is the Boltzman constant. Remarkably, this relationship is consistent with the $\chi''$-peak temperature change between the two ensembles with $<D>=20$ nm to $<D>=13$ nm, which strongly suggests that the magnetic behavior observed below $T_N$ stems form the superparamagnetic relaxation of the ferrimagnetic nanoparticle ensemble and not from a collective freezing of surface spins.

To confirm the superparamagnetic origin of the magnetic anomaly observed below $T_N$, we carried out a detailed analysis of the frequency and particle-size dependence of the $\chi''$-peak temperature T. We found that the relative variation of T per frequency decade $\dfrac{\Delta T}{T \cdot \log(f)}$ falls within the typical range of values expected for superparamagnetic blocking, which is one order of magnitude greater than that of its spin-glass-transition counterpart. We also observed that a Vogel-Fulcher law $\tau = \tau_0 \exp[\Delta E_B / k_B (T - T_0)]$ accurately describes the shift of T with the observation time $\tau = 1/2\pi f$, yielding a barrier to magnetization reversal $E_B$, a time constant $\tau_0$, and an interparticle-interaction strength parameter $T_0$, whose values are consistent with the superparamagnetic relaxation of a slightly interacting system of nanoparticles. Finally, we demonstrate that even a slight change in ensemble's average size leads to a measurable shift in the $\chi''$-peak temperature (recorded at a given frequency), and that the magnitude of the shift is in quantitative agreement with the nanoparticle-size dependence of the superparamagnetic blocking temperature.

**EXPERIMENTAL DETAILS**

Mn$_3$O$_4$ nanocrystals were synthesized using a co-precipitation technique whereby sodium hydroxide was added dropwise to a MnCl$_2$·4H$_2$O solution in deionized water. The size of the precipitated nanoparticles was controlled by varying the molarity of the manganese chloride solution. After the slow addition of NaOH the samples were heated to 90ºC for 30 minutes to convert the manganese hydroxide to the Mn$_3$O$_4$ nanomaterial. Finally, the samples were centrifuged at 3,000 rpm for 10 minutes, rinsed with deionized water 3 times, and allowed to air dry for 24 hours.

The crystal structure and impurity-free nature of the Mn$_3$O$_4$ ensembles were confirmed via laboratory x-ray powder diffraction measurements carried out using a Siemens D5000 diffractometer (wavelength $\lambda$ = 1.5406 Å) equipped with a Braun position sensitive detector. The sample was loaded in a flat-plate holder and diffraction patterns were collected in the reflectivity geometry for d-spacing values between 1.5 Å and 3.5 Å, or, equivalently, over the 20º-60º detector angle (2$\theta$) range. Data collection time for each diffraction pattern was approximately 60 minutes. The process was repeated several times to ensure the reproducibility of the results. No significant difference was observed between experimental runs.

The average size and size distribution of the Mn$_3$O$_4$ nanoparticles was determined from transmission electron microscopy (TEM) measurements conducted using a Hitachi H-9500, high-resolution microscope operating at 300kV, utilizing a goniometer-tilt stage and fitted with a CCD digital imaging camera. The samples were suspended in a pyridine solution and solution drops deposited onto 3mm, silicon monoxide coated, 200 mesh copper TEM grids, while a second grid was placed on the dried Mn$_3$O$_4$ particle deposit to form a sandwich. The grid sandwich was then placed in the TEM and images recorded in both bright and dark field using direct magnifications ranging from 20,000x to 700,000x; or digitized from 200kX to 2,000 kX. In addition, selected-area electron diffraction patterns were taken to confirm the x-ray diffraction results and to obtain systematic dark-field images and high-resolution lattice images. Nanoparticle sizes were determined using a random grid overlay on enlarged, dark-field images and measured for particles falling upon the grid lines. Finally, particle sizes were plotted in histograms and the average particle diameter <D> determined. The same magnifications (100,000x) and statistical measuring areas were maintained for all particle measurements, with essentially the same sample size (or number of measurements, N).

Ac magnetic susceptibility measurements were carried out using a Quantum Design® Physical Properties Measurement System (PPMS). Approximately 25 mg of sample was loaded into a polycarbonate capsule, attached to the sample rod, and lowered into the cryostat of the PPMS. The in-phase and the out-of-phase components of the ac susceptibility were recorded over the temperature range from 3 K to 50 K upon heating at different frequencies between 100 Hz and 10 kHz. For all measurements the amplitude of the applied alternating magnetic field was 3 Oe.

**RESULTS AND DISCUSSION**

Figure 1 shows the x- ray diffraction (XRD) pattern from a $Mn_3O_4$ nanoparticle ensemble. The open symbols represent the observed intensity recorded for different d-spacing values between 1.5 and 3.5 Å. The solid line is a full profile (Le Bail) fit [22] to the data, the vertical bars indicate the d-spacing positions of the Bragg reflections, and the lower trace is the difference curve between the observed and the calculated intensities. The fit confirms that the sample consists of a single nanocrystalline $Mn_3O_4$ phase with tetragonal ($I4_1/amd$) symmetry and lattice constants a = 5.76 Å and c = 9.44 Å.

Transmission electron microscopy (TEM) was used to accurately determine the average size and size distribution of the nanoparticles in the two ensembles used in this study. Figure 2 shows (a) bright-field and (c) dark-field images of one of the two samples. The dark-field image was obtained by positioning the objective aperture over a region corresponding to combined {011}, {112}, and {020} diffraction spots (circled area in Fig. 2 (b)), which simultaneously showed differentiated, crystalline nanoparticles whose sizes (diameters) could be measured from random grid line intersections. These measurements were plotted in histograms (Fig. 3 (a) and (b)) and the average nanoparticle diameters <D> fore the two ensembles were determined to be 13 and 16 nm. Figure 2 (d) illustrates (020) lattice plane images for several overlapping $Mn_3O_4$ particles which create rotation Moiré fringes. The long arrows indicate traces of (020) planes in two crystals M1 and M2 oriented at an angle of about 16° with respect to one another. The d-spacing of the (020) planes is $d_{020} \cong 2.9$ Å, while the Moiré fringe spacing is 8.7 Å and 10.4 Å in M1 and M2 nanocrystals, respectively. These crystalline areas are best viewed by sighting along the long arrows or parallel to the Moiré fringes. The short arrows indicate the edges of nanoparticle crystal M2 - it can be observed that the fringes extend to the nanoparticle edges, indicative of crystallinity extending to the particle surfaces with no indication of an amorphous surface layer.

The temperature dependence of the out-of-phase ac-susceptibility $\chi''$ measured within the 3 K – 50 K range on the <D> = 13 nm $Mn_3O_4$ nanoparticle ensemble is shown in Fig. 4. The five datasets were collected at different frequencies of the driving magnetic field: f=100 Hz (open circles), f=300 Hz (filled circles), f=1 kHz (open triangles), f=3 kHz (filled triangles), and f=10 kHz (open squares). The first feature revealed by the $\chi''$ vs. T curves is a pronounced frequency-independent peak at about 40 K, which corresponds to the cooling-induced transition of $Mn_3O_4$ to its ferrimagnetic Yafet-Kittel phase [1,2]. Below this temperature, our data does not show any evidence of the complex magnetic ordering previously observed in bulk $Mn_3O_4$ [1-4]. Instead, we found a weak and relatively broad $\chi''$ peak at about 11 K. More detailed measurements of this feature (shown in the inset) revealed a well-defined frequency dependence of the peak temperature, which slightly increases with increasing f. Although a similar magnetic behavior was recently observed below the paramagnetic-to-ferrimagnetic-transition temperature in larger $Mn_3O_4$ nanoparticles (<D>=20 nm) [5], its microscopic origin has not been yet clarified. The authors of Ref. 5 hint at the possibility that this frequency dependent magnetic anomaly, observed by them near 31 K, might be due to the collective spin freezing of a surface layer that surrounds the ferrimagnetic core on the nanoparticles. Yet, in view of our above-presented findings, the 20 K difference in the peak position between the two studies (in the context of the increase of the nanoparticle average size) appears not to support a spin glass like transition origin, since the spin freezing temperature is not expected to change significantly with <D>. Instead, the fact that the peak temperature T roughly changes as $<D>^3$ strongly suggests a superparamagnetic relaxation origin, as the blocking temperature $T_B$ of an ideal magnetic nanoparticle ensemble is indeed proportional to the cube of the nanoparticle size according to:

$$T_B \propto \frac{K<D>^3}{k_B} \qquad (1),$$

where K is the magnetic anisotropy assumed to be uniaxial and $k_B$ is the Boltzman constant.

In this scenario, cooling down $Mn_3O_4$ nanoparticles below the Néel temperature ($T_N \cong 41K$) leads to a ferromagnetic-to-paramagnetic transition in the nanoparticles' material. However, for small enough nanoparticles, the nanoparticle ensemble still behaves (super)paramagnetically at temperatures below $T_N$. In our <D>=13 nm ensemble, for example, all nanoparticles of sizes smaller than 20 nm (i.e the majority of the ensemble) are in their unblocked (superparamagnetic) state even at 30 K. Upon further cooling, an increasing number of nanoparticles become blocked with the average sized (13 nm) ones blocking at $T_B \sim 11$ K. Even lower temperatures are required for the majority of the ensemble to block. Below, we present evidence for the superparamagnetic nature of the behavior of $Mn_3O_4$ nanoparticle ensembles below $T_N$.

The low-temperature-peak positions T were accurately determined at each frequency f (or, equivalently, observation time $1/2\pi f$) from polynomial fits to the $\chi''$ vs. T data (solid lines in the inset to Fig. 4). These values were first used to calculate the relative variation of the peak temperature per frequency decade. We found $\frac{\Delta T}{T \cdot \log(f)} \cong 0.11$, a value that is one order of magnitude larger than that expected for spin freezing [8,13], but within the range of values commonly observed in superparamagnetic systems [15]. To further test the origin of the system's low-temperature dynamic behavior, we analyzed the observation time dependence of the peak temperature on the basis of conventional dynamic scaling theory which holds that the relaxation time, $\tau$, of a system diverges as a power law within the correlation length, $\xi$, such that $\tau = \tau_0 \xi^z$. Here $\tau_0$ is a characteristic time constant (related to the attempt frequency by the relation $\tau_0 = 1/2\pi f_0$) and z is a dynamic scaling exponent. In addition, according to the static scaling hypothesis, $\xi = [(T/T_f)-1]^\nu$, where $T_f$ is the critical freezing temperature and $\nu$ is a critical exponent. One eventually finds:

$$\tau = \tau_o \left[ \frac{T}{T_f} - 1 \right]^{-z\nu} \quad (2)$$

where, in our case, $\tau$ is the observation time (as the relaxation time of the system is equal to the observation time at the $\chi''$ peak temperature T). Figure 5 shows a least-squares fit of Eq. (2) (solid line) to the observed $\tau$ vs. T dependence (solid symbols). The fit converges to low residuals yielding parameters $\tau_0 \sim 10^{-7}$ s, $z\nu = 4.6$, and $T_f = 9.1$ K. Significantly, however, the values for the time constant $\tau_0$ and for the exponent $z\nu$ are outside of the range of values expected for spin-glass like transitions (i.e. and $\tau_0$ between $10^{-11}$ s and $10^{-13}$ s and $z\nu$ between 8 and 10 [23]). This further indicates that surface spin freezing is not likely to be the microscopic origin of the low-temperature magnetic behavior observed in $Mn_3O_4$ nanoparticle ensembles.

The superparamagnetic relaxation of an ideal system of noninteracting, single-domain, and monodisperse magnetic nanoparticles is described by the Néel-Brown equation $\tau = \tau_o \cdot \exp\left[\frac{E_B}{k_B T}\right]$, which predicts how rapidly the magnetic moment of a single particle flips along an easy axis by thermal activation [6,7]. Here $\tau$ is the relaxation time at a given temperature, $E_B$ the energy barrier to superspin reversal, $k_B$ the Boltzman constant, and $\tau_0$ a time constant. To investigate if superparamagnetic blocking is indeed responsible for the magnetic behavior of the $Mn_3O_4$ nanoparticles below $T_N$, we first made an attempt at fitting the Néel-Brown equation to our observed $\tau$ vs. T dependence obtained from frequency dependent ac-susceptibility measurements. The best fit describes this dependence well, but yields an unphysically short time constant $\tau_0 \cong 10^{-14}$s (the shortest timescale in magnetism is the spin flip time of a single atom $\tau_s \sim 10^{-13}$ s). As previously demonstrated [12], this indicates that the magnetic nanoparticle system is not ideal, and interparticle interactions play a non negligible role; in this case in a Vogel-Fulcher law:

$$\tau = \tau_o \cdot \exp\left[\frac{E_B}{k_B(T-T_o)}\right] \qquad (3)$$

has been proven to describe the system's relaxation [19,20] and to provide information about the strength of the interparticle interactions through the value of the additional parameter $T_o$. As shown in Fig. 6, we found that the measured $\tau$ vs. T dependence (solid symbols) is excellently described by Eq. (3); the best fit (solid line) yields $E_B/k_B$= 87 K, $\tau_0 \sim 10^{-10}$ s and $T_0$ = 4.8 K, all within the range expected for a slightly interacting superparamagnetic system.

Another signature of superparamagnetic relaxation is that the blocking temperature measured at a given frequency (which, in our case, is determined as the $\chi''$-peak temperature T) is sensitive to small changes in the ensemble's average size <D>. Moreover, if interactions are present in the system, and the parameter $T_0$ that describes the strength of such interactions is known, a quantitative relationship between T and <D> can easily be derived. Indeed, using Eq. (3) and the fact that the barrier to magnetization reversal is proportional to the magnetic anisotropy constant and the cube of the average nanoparticle size (i.e. $E_B \propto K<D>^3$) on finds:

$$(T - T_0) \propto \langle D \rangle^3. \qquad (4)$$

Figure 7 shows the temperature dependence of the out-of phase susceptibility measured at f=100Hz on two $Mn_3O_4$ nanoparticle ensembles: one with average nanoparticle size $\langle D_1 \rangle$=13nm (solid symbols) and another with slightly larger nanoparticles $\langle D_2 \rangle$=16nm (open symbols). We first note that the $\chi''$-peak temperature shifts from $T_1$=10.3 K to $T_2$=14.8 K upon the increase of the nanoparticle average size. This is clearly consistent with a superparamagnetic blocking rather than a spin freezing origin of the observed $\chi''$ vs. T behavior. Even more importantly, the magnitude of the above-mentioned temperature shift is in excellent quantitative agreement with the behavior of the superparamagnetic blocking temperature upon the nanoparticle-size variation predicted by Eq. (4). Indeed, using $T_0$ = 4.8 K, we find $\frac{T_1 - T_0}{T_2 - T_0} \approx \left( \frac{\langle D_1 \rangle}{\langle D_2 \rangle} \right)^3$. This is highly significant, as the 4.8 K value of the interaction strength parameter has been obtained via a Volger-Fulcher law fit to data from frequency-resolved $\chi''$ observations, i.e. in a totally independent measurement from the one used to yield the T vs. <D> dependence. We believe this represents strong evidence for the superparamagnetic nature of the behavior of fine $Mn_3O_4$ nanoparticles below the Néel temperature of the title material

**SUMMARY**

We have investigated the magnetic behavior of $Mn_3O_4$ nanoparticle ensembles at temperatures below the paramagnetic-to-ferrimagnetic transition ($T_N \cong 41$ K) using ac-susceptibility measurements and transmission electron microscopy. Our data confirm the suppression of the low-temperature complex magnetic order exhibited in bulk $Mn_3O_4$.

We found a frequency-dependent peak in the out-of-phase component of the ac magnetic susceptibility $\chi''$ at a temperature near 11 K, a type of magnetic anomaly previously observed in Ref [5] and tentatively associated with surface spin effects. Our analysis of the frequency and average-particle-size dependence of the $\chi''$ vs. T data demonstrates that, in fact, superparamagnetic blocking is responsible for this low-temperature magnetic behavior. We found that the relative variation of the peak temperature per frequency decade $\Delta T/T\Delta\log(f)$ is one order of magnitude larger than the values typically observed for collective spin freezing, but within the range expected for superparamagnetic systems. In addition, attempts to fit the frequency (observation time) dependence of the $\chi''$ vs. T peak by a power law according to the dynamic scaling theory led to parameter values well outside of the predicted range for glassy transitions. Yet, this dependence is very well described by a Vogel-Fulcher law, which is known describe the superparamagnetic relaxation of ensembles of slightly interacting magnetic nanoparticles. Finally we demonstrated that the shift of the $\chi''$ peak temperature upon the modification of the ensemble's average-size is consistent with the expected variation of the superparamagnetic blocking temperature of a nanoparticle system.

**REFERENCES**

[1] K. Dwight and N. Menyuk, *Phys. Rev.* **119**, 1470 (1960).

[2] G.B. Jensen and O.V. Nelson, *J. Phys. C* **7**, 409 (1974).

[3] T. Suzuki and T. Katsufuji, *Phys. Rev. B* **77**, 220402(R) (2008).

[4] R. Tackett, G. Lawes, B.C. Melot, M. Grossman, E.S. Toberer, and R. Sheshadri, *Phys. Rev. B* **76**, 024409 (2007).

[5] R. Regmi, R. Tackett, and G. Lawes, *J. Magn. Magn. Mater*. **321**, 2296 (2009).

[6] L. Néel, *Ann. Geophys*. **5**, 1677 (1949).

[7] W.F. Brown, *Phys. Rev.* **130**, 1677 (1963).

[8] R.H. Kodama, A.E. Berkowitz, E.J. McNiff Jr., and S. Foner, *Phys. Rev. Lett*. **77**, 394 (1996).

[9] B. Martinez, X. Obradors, L.I. Ballcels, A. Rouanet, and C. Monty, *Phys. Rev. Lett.* **80**, 181 (1998).

[10] S. Sahoo, O. Petracic, Ch. Binek, W. Kleenman, J.B. Sousa, S. Cardoso, and P.P. Freitas, *Phys. Rev. B* **65**, 134406 (2002).

[11] S. Sahoo, O. Petracic, W. Kleenman, S. Stappert, G. Dumpich, P. Nordblad, S. Cardoso, and P.P. Freitas, *Appl. Phys. Lett*. **82**, 4116 (2003).

[12] R.J. Tackett, A.W. Bhuiya, and C.E. Botez, *Nanotechnology* **20**, 445705 (2009).

[13] J.L. Tholence, A. Benoit, A. Mauger, M. Escorne, and R. Triboulet, *Solid State Commun.* **49**, 417 (1984).

[14] J. Ferre, J. Rajchenbach, and H. Maletta, *J. Appl. Phys*. **52**, 1829 (1981).

[15] T. Mori, H. Mamiya, *Phys. Rev. B* **68**, 214422 (2003).

[16] Y. Zhou, C. Rigaux, A. Mycielski, M. Menant, and N. Bontemps, *Phys. Rev. B* **58**, 8111 (1989).

[17] P.M. Shand, A.D. Christianson, T.M. Pekarek, L.S. Martinson, J.W. Schweitzer, I. Miotkowski, and B.C. Crooker, *Phys. Rev. B* **58**, 1286 (1998).

[18] J.T. Elizalde Galindo, A.H. Adair, C.E. Botez, V. Corral Flores, D. Bueno Baques, L. Fuentes Cobas and J.A. Matutues-Aquino, *Appl. Phys. A* **87**, 743 (2007).

[19] H. Vogel, Z. Phys. 22, 645 (1921). G.S. Fulcher, J. Am. Ceram. Soc. 8, 339 (1925).

[20] J. Zhang, C. Boyd, and W. Luo, *Phys. Rev. Lett*. **77**, 390 (1996).

[21] D.L. Leslie-Pelecky and R.D. Rieke, *Chem. Mater.* **8**, 1770 (1996).

[22] A. Le Bail, H. Duroy, and J.L. Fourquet, *Mat. Res Bull.* **23**, 447 (1988).

[23] P. Granberg, J. Mattson, P. Nordblad, L. Lundgren, R. Stubi, J. Bass, D.L. Leslie-Pelecky, and J.A.Cowen, *Phys. Rev. B* **44**, 4414 (1991).

**FIGURE CAPTIONS**

Figure 1   X-ray powder diffraction pattern measured on a $Mn_3O_4$ nanoparticle ensemble (open symbols). The solid line is a full-profile (Le Bail) fit which confirms that the sample consists of a single nanocrystalline $Mn_3O_4$ phase with tetragonal ($I4_1/amd$) symmetry and lattice constants a = 5.76 Å and c = 9.44 Å.

Figure 2   TEM of an aggregation of $Mn_3O_4$ nanoparticles: (a) bright field TEM micrograph, (b) electron diffraction pattern, (c) dark-field TEM micrograph, and (d) high resolution TEM micrograph showing lattice planes (large arrows) for overlapping nanoparticles that create Moiré fringes at M1 and M2. The small arrows indicate that the particles show no amorphous surface layer.

Figure 3   Histograms of the particle size distribution for the two $Mn_3O_4$ nanoparticle ensembles used in this study. The histograms show the ensembles' polydispersity and yield average diameters of (a) 13 nm and (b) 16 nm.

Figure 4   Temperature dependence of the out-of-phase magnetic susceptibility $\chi''$ measured on a $<D>=13$ nm average particle size $Mn_3O_4$ nanoparticle ensemble at five different frequencies: f=100 Hz (open circles), f=300 Hz (filled circles), f=1 kHz (open triangles), f=3 kHz (filled triangles), and f=10 kHz (open squares). The solid lines in the inset are fits to polynomial functions that allow a precise determination of the $\chi''$ vs. T peak position for each frequency.

Figure 5   Best fit of Eq. (2) (solid line) to the observed frequency / observation time dependence of the $\chi''$ peak temperature (solid symbols).

Figure 6   Vogel-Fulcher law best fit (solid line) to the observed $\tau$ vs. T dependence (solid symbols). The fit yields $E_B/k_B = 87$ K, $\tau_0 \sim 10^{-10}$ s and $T_0 = 4.8$ K.

Figure 7   Temperature dependence of the out-of-phase magnetic susceptibility $\chi''$ measured at f = 100 Hz on $Mn_3O_4$ nanoparticle ensembles of average diameters $<D_1>$ = 13 nm (solid symbols) and $<D_2>$ = 16 nm (empty symbols). The $\chi''$-peak temperature shifts from $T_1=10.3$ K to $T_2 =14.8$ K upon the increase of the nanoparticle average size.

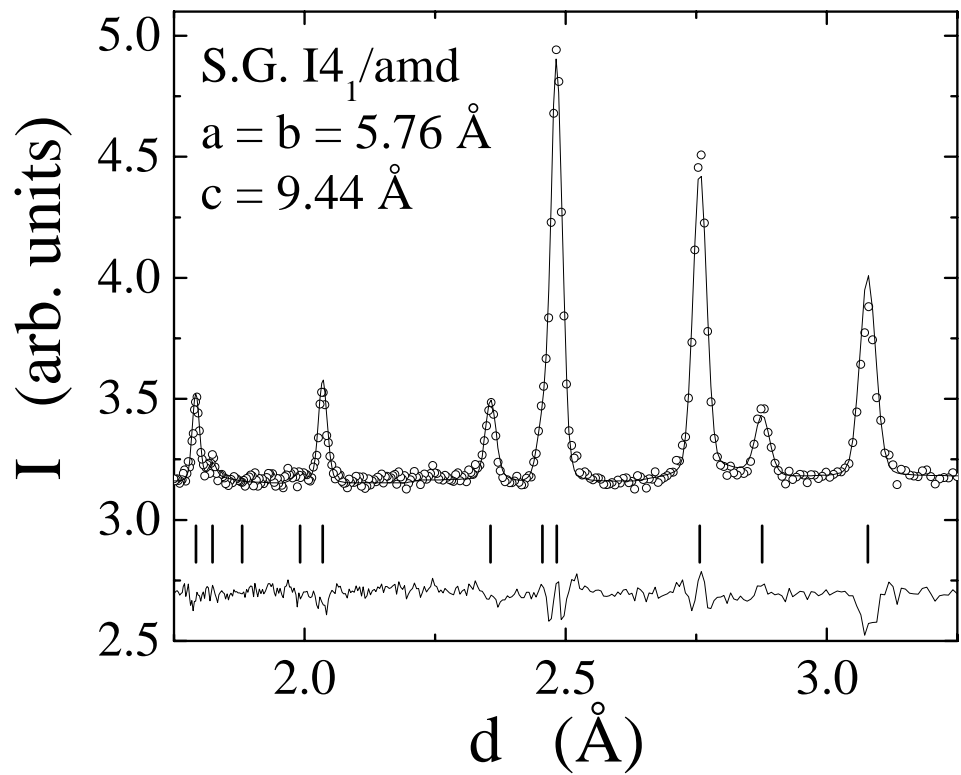

Figure 1　　　　　　　　　　　　　　　　　　　　　　　　　　　　Tackett et al.

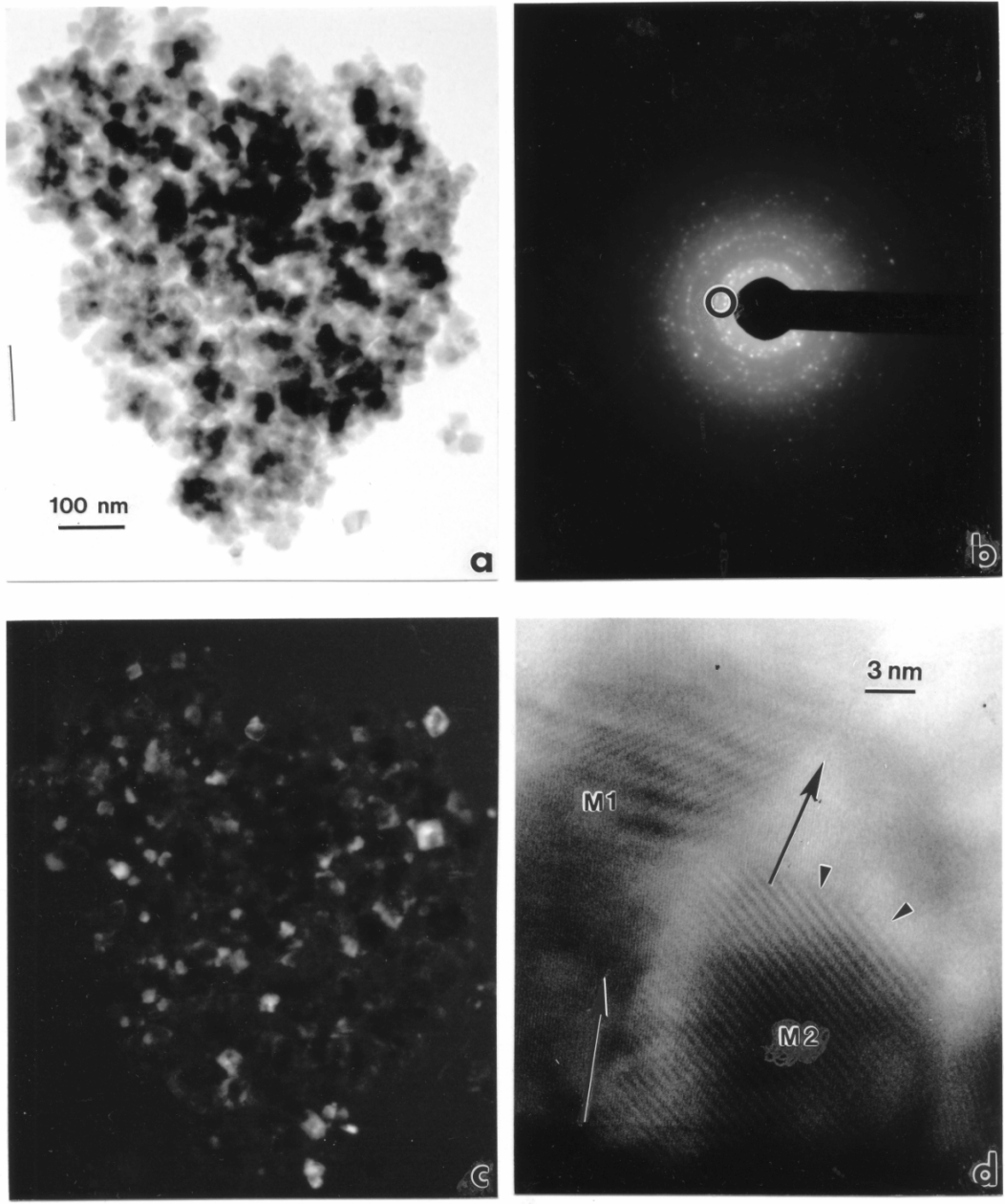

Figure 2               Tackett et al.

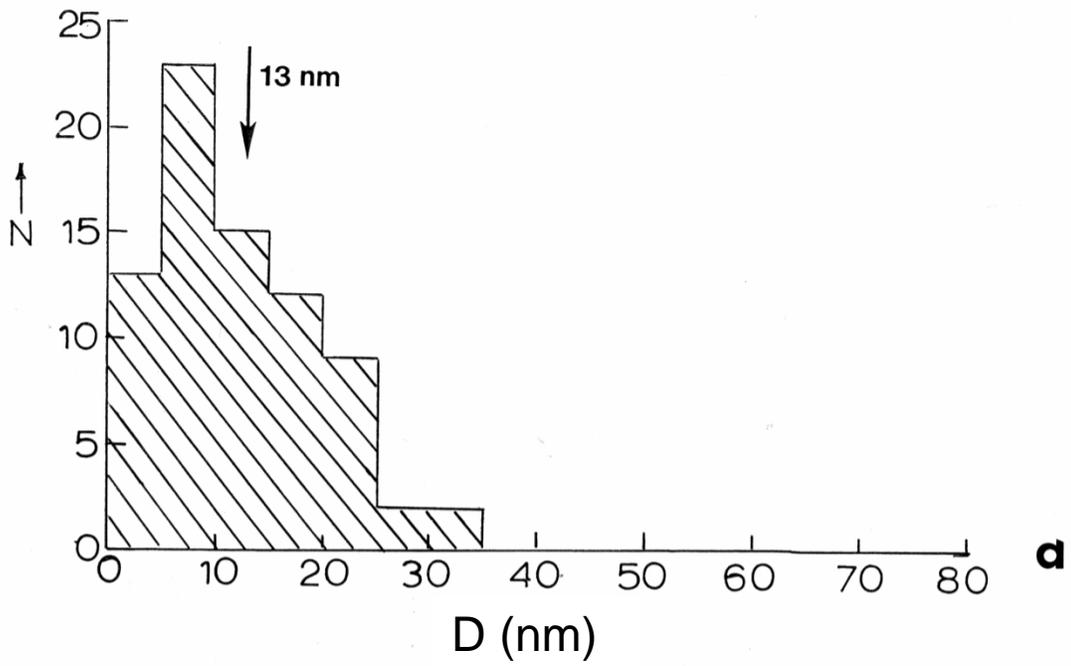

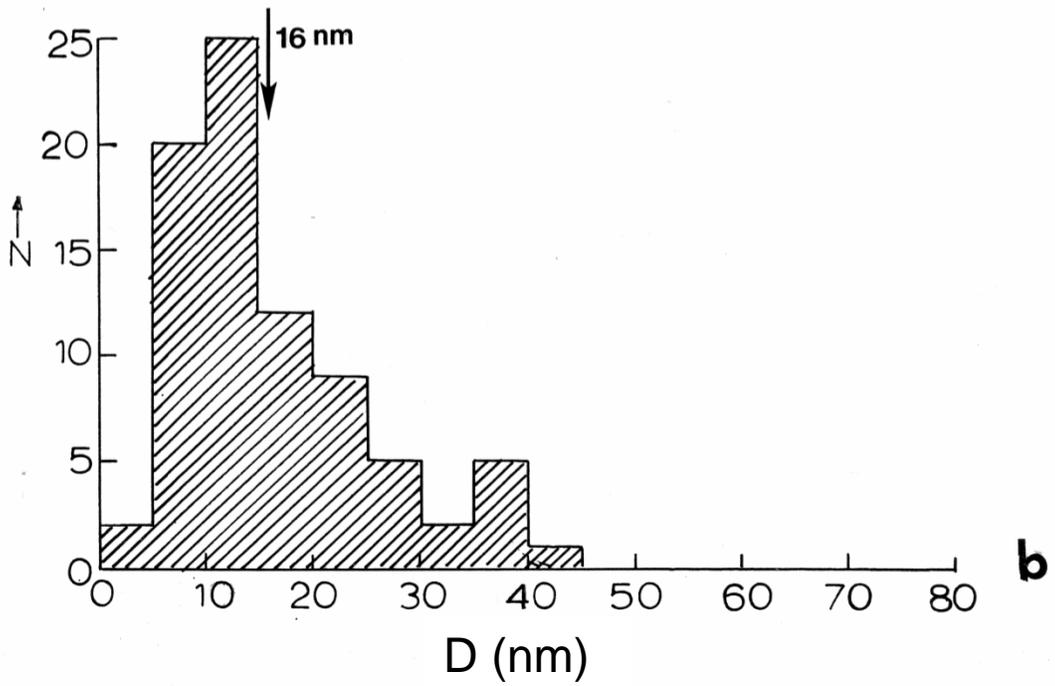

Figure 3        Tackett et al.

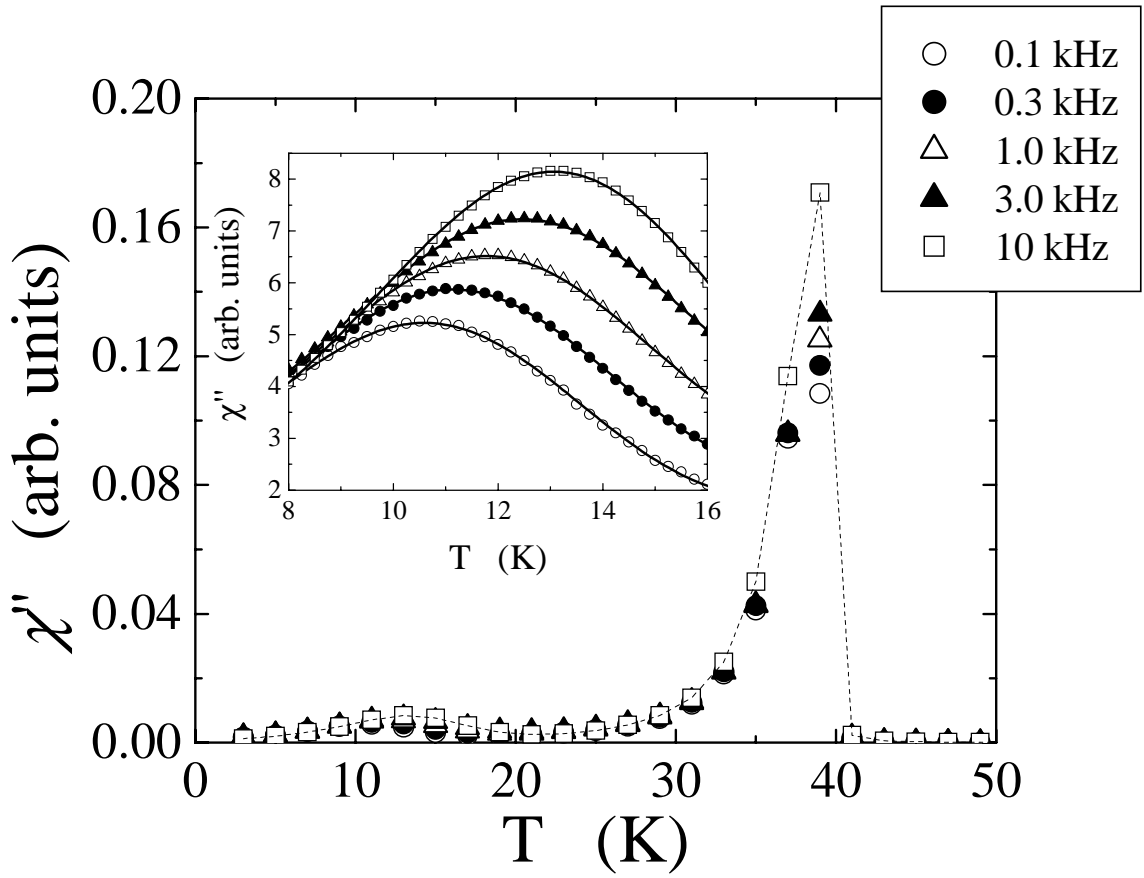

Figure 4 Tackett et al.

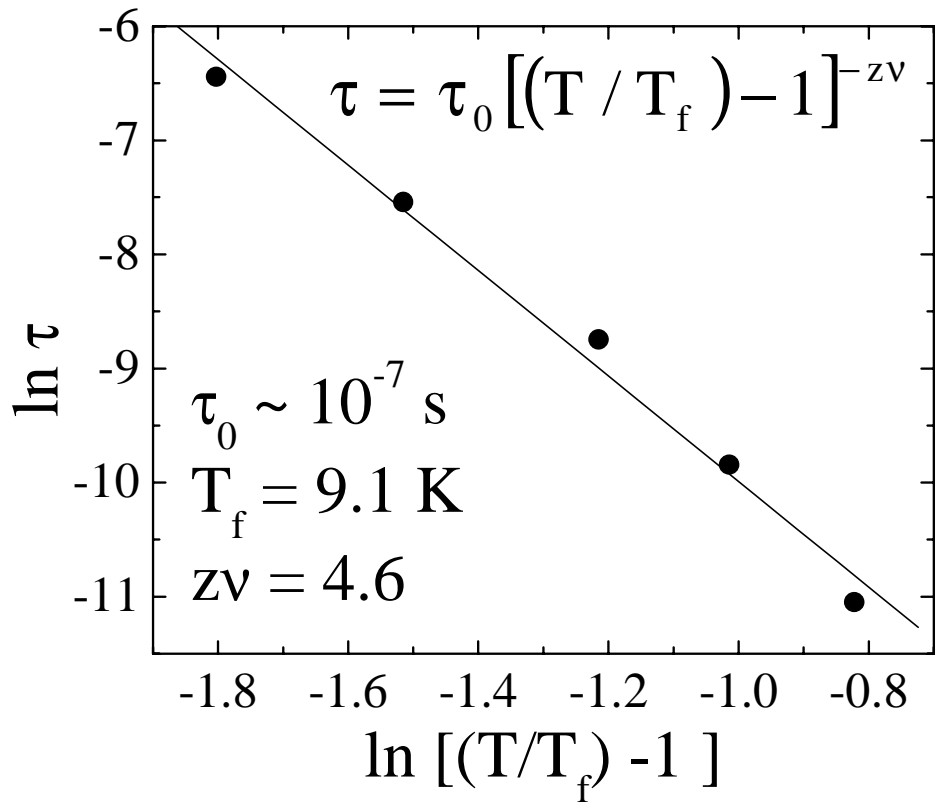

Figure 5 Tackett et al.

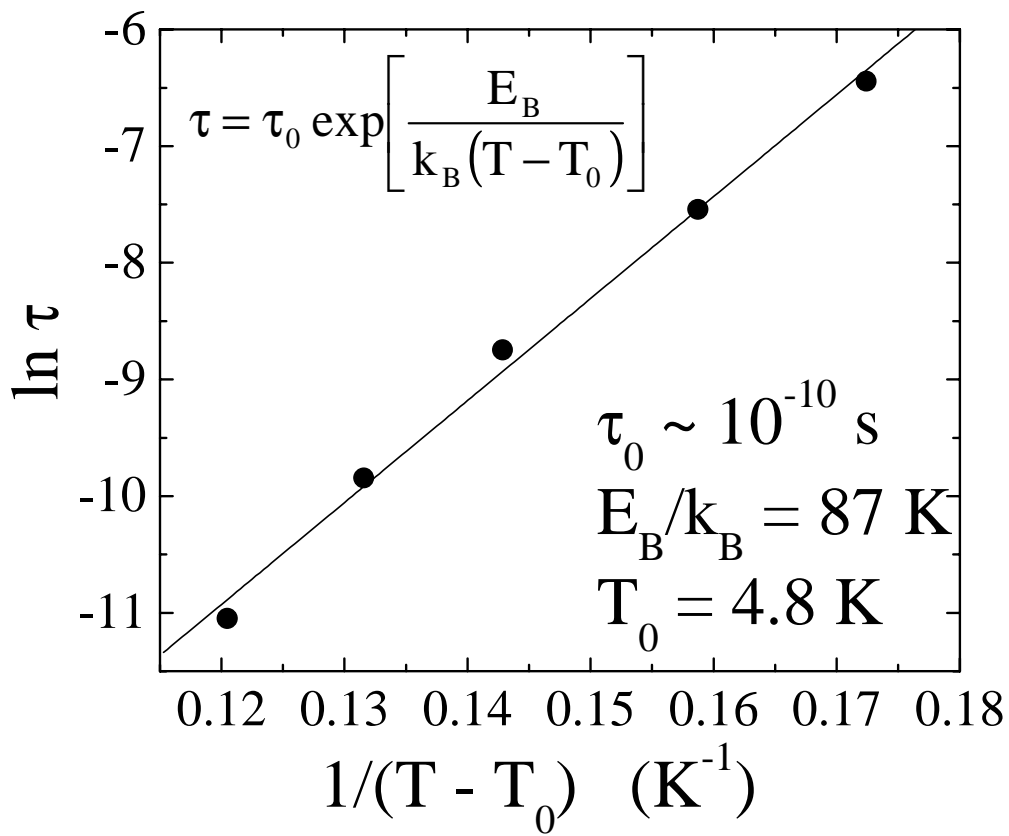

Figure 6 Tackett et al.

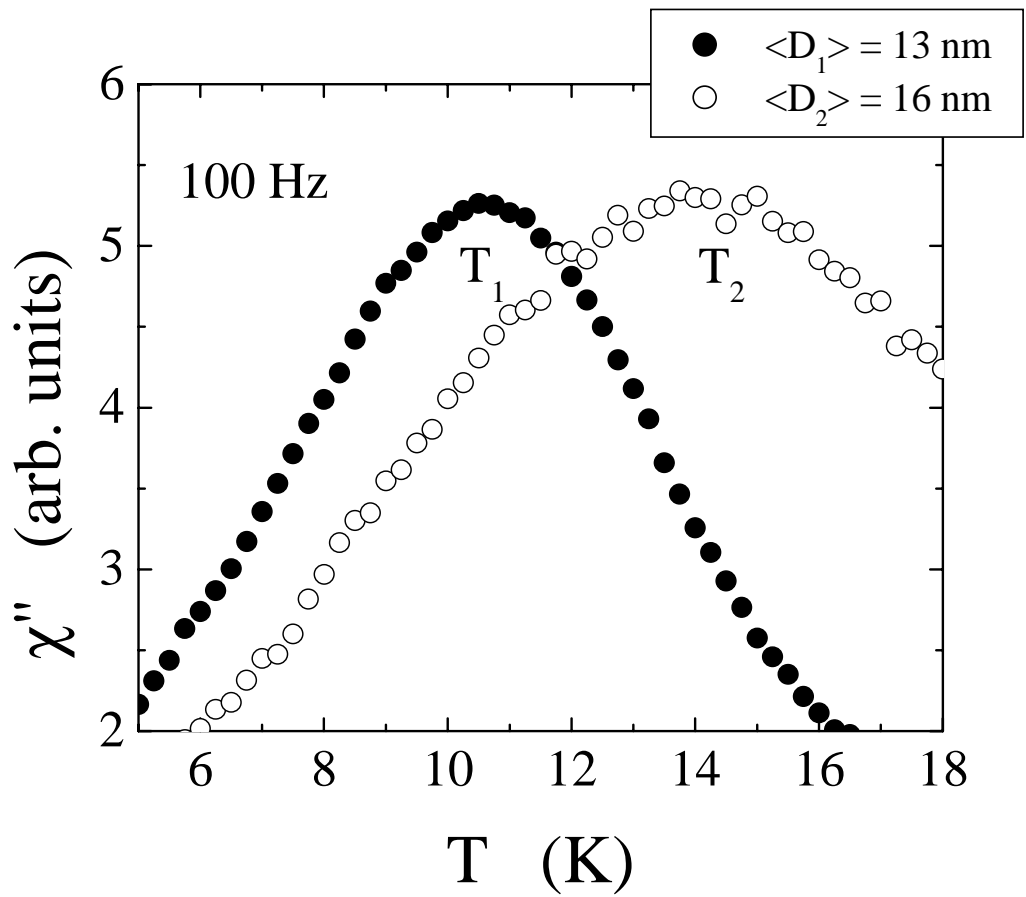

Figure 7 — Tackett et al.